\documentclass[final,1p,times]{elsarticle}

\usepackage{graphicx}
\usepackage{amssymb}
\usepackage{hyperref}

\usepackage{lineno}

\makeatletter
\def\ps@pprintTitle{%
 \let\@oddhead\@empty
 \let\@evenhead\@empty
 \def\@oddfoot{}%
 \let\@evenfoot\@oddfoot}
\makeatother
\begin{document}

\begin{frontmatter}


\title{Using Macroeconomic Forecasts to Improve Mean Reverting Trading Strategies}



\author{Yash Sharma}

\begin{abstract}
A large class of trading strategies focus on opportunities offered by the yield curve. In particular, a set of yield curve trading strategies are based on the view that the yield curve mean-reverts. Based on these strategies' positive performance, a multiple pairs trading strategy on major currency pairs was implemented. To improve the algorithm's performance, machine learning forecasts of a series of pertinent macroeconomic variables were factored in, by optimizing the weights of the trading signals. This resulted in a clear improvement in the APR over the evaluation period, demonstrating that macroeconomic indicators, not only technical indicators, should be considered in trading strategies.
\end{abstract}

\end{frontmatter}

\section{Yield Curve}

The Yield Curve is a line that plots the interest rate, at a set point in time, of bonds having equal credit quality but differing maturity dates. The U.S. Treasury Yield Curve is used as a benchmark for other debt in the market, and can also be used to predict changes in economic output and growth. The shape of the yield curve gives an idea of future interest rate changes and economic activity.  An inverted yield curve is often a harbinger of recession. A positively sloped yield curve is often a harbinger of inflationary growth. These predictive properties make the yield curve quite applicable in devising successful trading strategies. \\

Many yield curve trading strategies are based on the conventional view that the yield curve mean-reverts to some historical norm. This market view is consistent with historical experience. For instance, U.S. Treasury bill rates, spreads and curvature all trade within tight, finite bounds. The interest rate term structures in other countries also exhibit similar patterns. This suggests that some form of mean-reversion mechanism is at work that prevents the yield curve from drifting to extreme levels or shapes over time. \\

In \cite{MeanRevert}, three classes of mean-reverting trading strategies were considered, focusing on three aspects of the yield curve: level, spread, and curvature. For each strategy, the holding period of a trade was fixed at one month, after which a new trade was initiated. The condition of cash neutrality was imposed, so that any excess cash was deposited at the 1-month tenor. Similarly, if additional funding was required, that was also carried out at the 1-month tenor. A 102-month training period was allowed in the construction of the unconditional yield curve, so that the calculation of the average payoff of each yield curve strategy starts from January 1973 to December 2000. \footnote{\textit{Profiting from Mean Reverting Yield Curve Trading Strategies}, 9}  \\

These mean-reverting trading strategies were compared to two benchmarks, investment in the Lehman Brothers U.S. Government Intermediate Bond Index, and cash-neutral investment in the S\&P index. \footnote{\textit{Profiting from Mean Reverting Yield Curve Trading Strategies}, 14} \\

The performance analysis indicated that a yield spread mean-reverting trading strategy performs remarkably well compared to the benchmarks. The monthly payoff is about 5.1 times that of the monthly payoff of the equity benchmark, hence outperforming an equity investment strategy, on a risk-adjusted basis. Furthermore, it outperformed the bond benchmark, with an average monthly payoff of about 5.9 times that of the benchmark. \footnote{\textit{Profiting from Mean Reverting Yield Curve Trading Strategies}, 16}\\

A paired-t test and the Diebold-Mariano statistical test (D-M test) \cite{Diebold} was conducted to test whether the strategy significantly outperforms the benchmarks. The tests were successful; even when transaction costs were accounted for, the yield spread mean-reverting strategy was still significantly more profitable than both of the benchmarks under all measures. \footnote{\textit{Profiting from Mean Reverting Yield Curve Trading Strategies}, 17-19}\\

As the yield curve is highly correlated with changes in economic activity, the yield spread mean reverting is a strong justification for trading upon the mean reversion of the spread, independent of the financial instrument used. Therefore, a Multiple Pairs trading strategy was implemented.

\section{Multiple Pairs Trading}

A Pairs Trade is a strategy based on securities involved in the pair having a mean-reverting nature. The goal is to match two trading vehicles, trading one long and the other short when the pair's price ratio diverges. The number of standard deviations the pair's price ratio diverges in order to initiate a trade is determined through historical data analysis. If the pair reverts to its mean trend, a profit is made. \\

In our implementation, currency pairs were used. Daily data from January 2008 to June 2014 for each of the major currency pairs was retrieved through Yahoo Finance. \\

\begin{figure}[h]
\centering\includegraphics[width=\textwidth,height=\textheight,keepaspectratio]{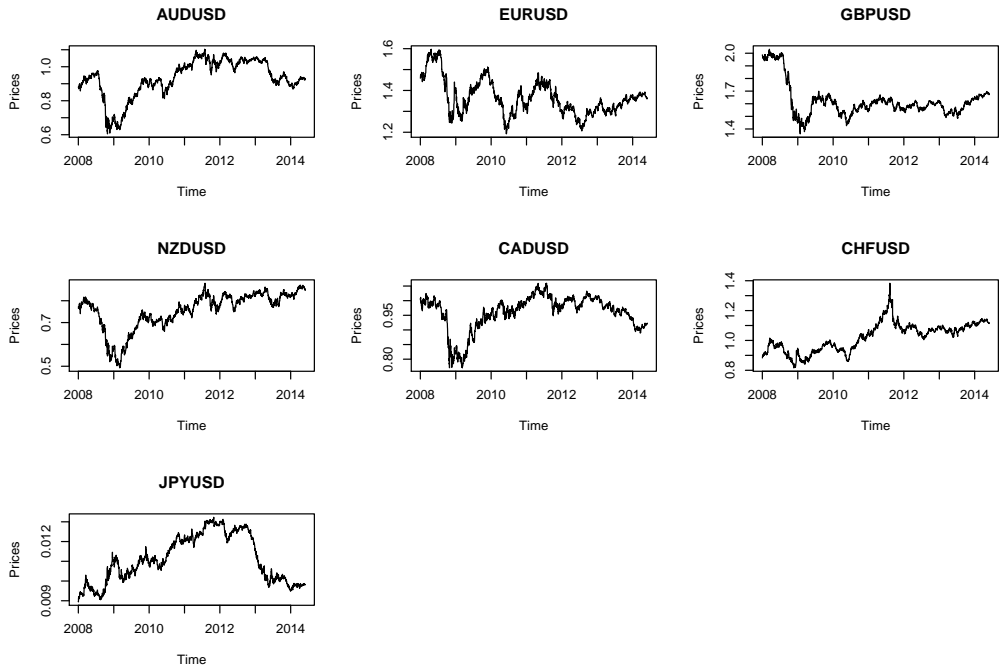}
\caption{Major Currency Pair Price Series}
\label{fig:PriceSeries}
\end{figure}

The currency pair price series were plotted, shown in Figure~\ref{fig:PriceSeries}. Clearly, the series do not look stationary. In order to perform a pairs trade, we desire a stationary pair, as the spread is fixed, and hence statistically the pair is mean reverting. \\

Consider a pair of non-stationary time series. If a particular linear combination of these series will lead to a stationary series, the pair of series are termed cointegrated. In order to find that particular combination, one needs to utilize tests for unit roots.

\subsection{Cointegration}

One would normally test for cointegration using the Augmented Dickey-Fuller test (ADF) \cite{ADF}. However, a big drawback of that test is that it is only capable of being applied to two separate time series. However, one can imagine a set of three or more financial assets that might share an underlying cointegrated relationship. Hence the Johansen test is used, which can be seen as a multivariate generalization of the ADF test. \cite{Johansen}The maximum number of currency pairs that can be combined in one relationship was set to 4. This gives the algorithm more flexibility, as all combinations of two, three, or four separate time series were tested for cointegration. \\

Before applying the Johansen test, or any unit root test for that matter, one must ensure that all the series are integrated of order 1. Technically, if two or more series are individually integrated, but some linear combination of them has a lower order of integration, then the series are said to be cointegrated. So, to guarantee that the cointegrated series is stationary, the time series need to be I(1). A price series is I(1) if the levels contain a unit root, meaning the price series is at least I(1), and the differenced prices are stationary, I(0), meaning the price series is not I(2). \\

The ADF test with the general regression equation was used, assuming the series has drift but no linear trend, in order to test for the existence of a unit root. Lags are included in the ADF formulation, allowing for higher-order autoregressive processes. The BIC information criterion is used to select the optimal lag length. \cite{BIC} This criterion is used for consistency reasons due to the large sample size. Though inefficient, the criterion delivers asymptotically correct results. Finally, the maximum lag length is chosen by the rule of thumb formula, proposed by Schwert, in \cite{Schwert}. The ADF test is performed on the levels and the differenced levels, and the test values were compared with the critical value at the 95\% confidence level to generate conclusions. \\

Each of the major currency pair price series were conclusively found to be I(1), and hence were used in the Johansen test. For the Johansen test, the VAR optimal solution lag length needs to be found. Again, the series was assumed to have drift but no linear trend, and the lag length was found by minimizing the SC information criterion for, again, the large sample size. Again, though inefficient, it provides asymptotically correct results. A lagged VAR is used in the Johansen procedure, so 1 is subtracted from the optimal VAR lag length. \\

The Johansen procedure was then initiated, again with the assumption of drift but no linear trend. The trace statistic was used, where the null hypothesis is that the number of cointegration vectors is less than or equal to the total number of vectors tested. Lastly, the longrun specification for error correction was used. \\

\begin{figure}[h]
\centering\includegraphics[width=\textwidth,height=\textheight,keepaspectratio]{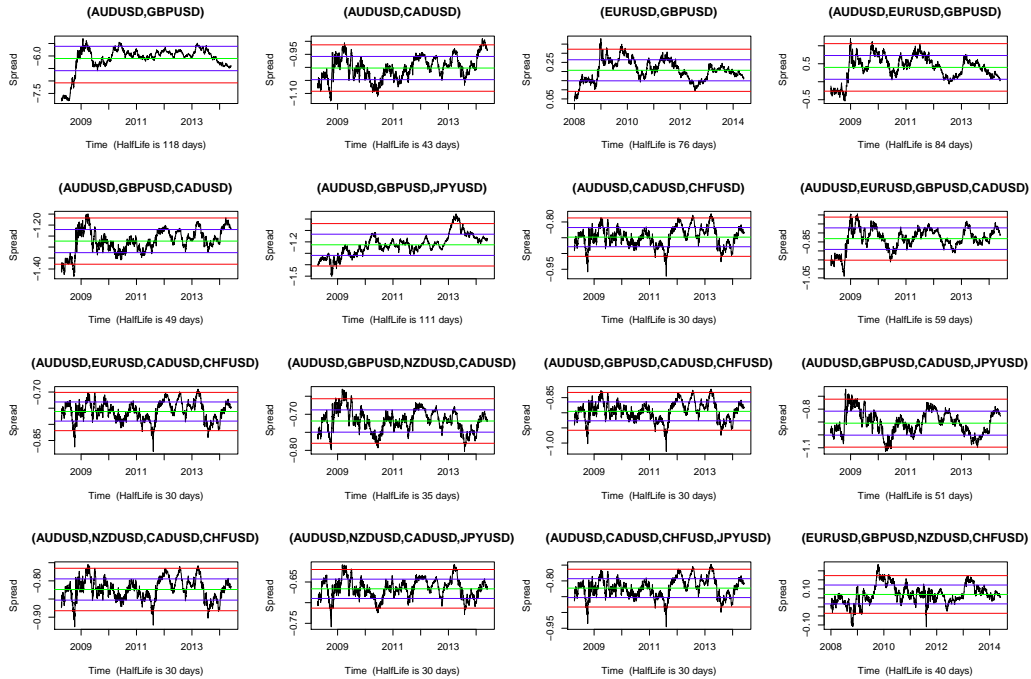}
\caption{Cointegrated Spreads}
\label{fig:CointegratedSpreads}
\end{figure}

91 combinations of time series were tested for cointegration, and 16 were found to be cointegrated. The spreads of the cointegrated portfolios were plotted, shown in  Figure~\ref{fig:CointegratedSpreads}. \\

As can be seen, the half life for each of the spreads was computed. Ignoring the drift and lagged difference terms, the differential form of the spread is equivalent to the Ornstein-Uhlenbeck stochastic process \cite{Brownian}. The differential form leads to an analytical solution for the expected value of the portfolio spread which, for a mean-reverting process with negative \(\lambda\), tells us that the expected value of the price decays exponentially at a half-life of \(-\log(2)/\lambda\). \(\lambda\), which measures the speed of an O-U process returning to its mean level, can be estimated from a linear regression of the daily change of the spread versus the spread itself. In order to perform this calculation, the spread and transaction costs are needed for the portfolio, and computing these costs will be discussed subsequently when the mean reversion strategy execution is described.

\subsection{Execution}

Finally, a mean reversion strategy was executed on each of the cointegrated spreads. Firstly, the spread and transaction costs that are incurred when buying/selling the cointegrated portfolio at the corresponding timestamp were computed. This computation depends on the individual price series of the instruments within the portfolio, and the optimal eigenvector associated with the results of the cointegration test. The eigenvectors generated from the Johansen test can be used as hedge ratios. The optimal eigenvector is the one that has the maximum eigenvalue, as it has the shortest half life.  \\

The spread is calculated by multiplying the hedge ratio values with their associated price values and a summation is made. If the portfolios are cointegrated, the resulting spread will be stationary. \\

The absolute value of the hedge ratio represents the number of units of each currency pair that we buy or sell. Hence, the total transaction costs can be computed with a summation on these amounts multiplied by the transaction costs per unit for each currency pair. \\

The mean and standard deviation of the spread were computed in order to calculate the z score, which simply represents the number of standard deviations separating the current price from the mean. If the  z score is positive, the current price of the security is above the mean. If the z score is negative, the current price of the security is below the mean. Hence, the z score is used to generate the mean reversion trading signals. \\

The entry z score was set to 1 and the exit z score was set to 0. These values set the extreme values, or thresholds, which when crossed by the signal, trigger trading orders. Entry into a long position is made if the z score is less than the negated entry score. Exit from a long position is made if the z score is greater than the negated exit score. Entry into a short position is made if the z score is greater than the entry score. Exit from a short position is made if the z score is less than the exit score. \\

With this, the long/short positions were set, yielding the number of units of the portfolio bought or sold at each timestamp. The USD capital allocation to buy the portfolio, specifically the USD capital invested in each currency pair at each timestamp, was found by computing the product between the hedge ratio matrix, which represents the hedge ratio of each currency pair at each timestamp, and the price matrix, which represents the price of each currency pair at each timestamp. \\

Using these computations, the P\&L of the strategy at each timestamp was computed. Transaction costs were subtracted from this value, and Return was found by dividing the resulting P\&L by the gross market value of the portfolio. From the return, the APR, Sharpe Ratio, and Maximum Drawdown of the strategy was computed. The spread (w/ half life), the standardized spread with color-coding indicating the long and short positions, the daily returns, and the cumulative returns were all plotted. Furthermore, the average and standard deviation of the daily returns, the APR, Sharpe Ratio, and Maximum Drawdown results were all reported. \\

\begin{figure}[h]
\centering\includegraphics[width=\textwidth,height=\textheight,keepaspectratio]{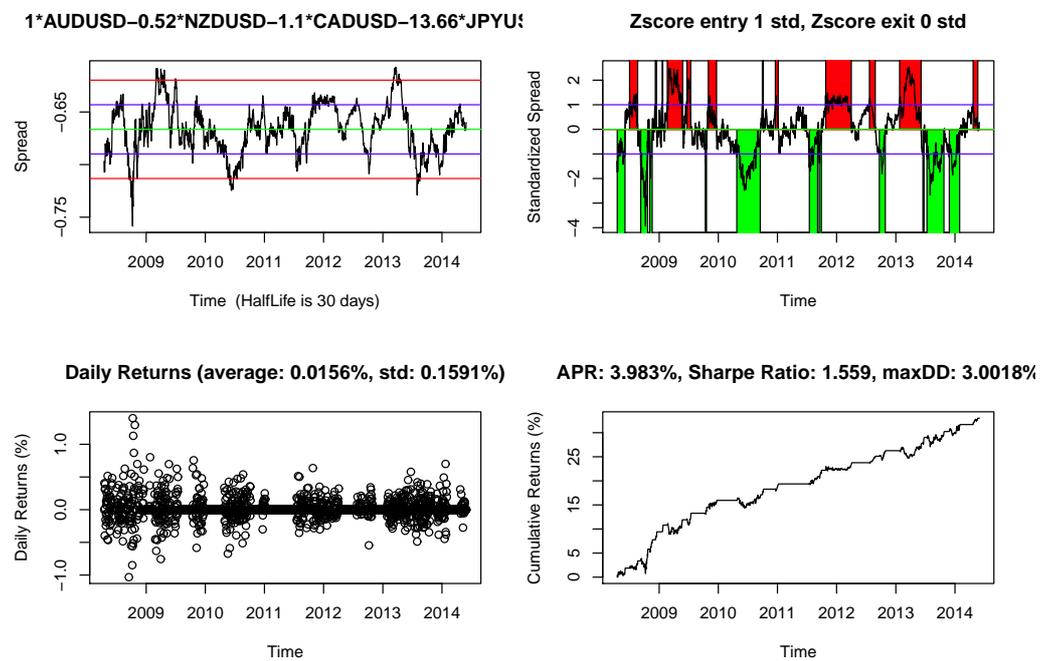}
\caption{Mean Reversion Results}
\label{fig:MeanReversionResults}
\end{figure}

For the forthcoming analysis, a cointegrated portfolio was randomly selected. The chosen portfolio consists of a combination of AUDUSD, CADUSD, NZDUSD, and JPYUSD currency pairs. The results generated after backtesting the strategy is shown in Figure~\ref{fig:MeanReversionResults}

\section{Macroeconomic Indicators}

Our goal was to use macroeconomic indicators to improve upon the implemented Multiple Pairs trading strategy. By generating buy/sell/hold signals based upon the forecasts of the chosen variables, and combining these signals with the mean reverting signal, we hoped to obtain improved performance. \\

There is much history with improving a Pairs Trading Strategy by factoring in additional variables, however the indicators chosen have typically been technical, and not macroeconomic. \\

In \cite{jiayu}, the authors' implemented Pairs Trading model was extended to take into account technical indicators. Rather than only considering the spread of the price, technical indicator movements were also considered. Technical indicators which were selected exhibited similar behaviors for both securities: Simple moving average (SMA), Weighted moving average (WMA), Money flow index (MFI), and Relative strength index (RSI). T-scores were computed for the price and the selected indicators, and a trained SVM was applied to the test dataset to make the final trading decisions. \\

This decision is typical, to not only use the price spread but other indicators intrinsic to the currency pair being traded. However, little exploration has been done for considering how macroeconomic movements reflecting the overall economy can improve trading performance. As currency pairs are representative of foreign exchange rates, with a downturn/upturn in the U.S. economy, a increase/decrease in the prices of currency pairs relative to the U.S. Dollar should take place. \\

Therefore, macroeconomic indicators which exhibit movements that are correlated with the strengthening/weakening of the U.S. Dollar were considered. 

\subsection{S\&P 500}

The Standard \& Poor's 500 Composite Stock Price Index was used. The S\&P Index Committee chooses the indexed stocks based upon market size, liquidity and industry group representation. Component companies are periodically replaced. Companies are most often removed because of a merger with another company, financial operating failure or restructuring. Prospective companies are placed in an index "replacement pool" and vacancies are filled from that pool.\\

The index is designed to measure changes in the stock prices of component companies. It is used as a measure of the nation's stock of capital, as well as a gauge of future business and consumer confidence levels. With that growth, the U.S. Dollar should strengthen. 

\subsection{Federal Funds}

The effective Federal Funds Rate was also used. It is the interest rate at which a depository institution lends funds maintained at the Federal Reserve to another depository institution overnight. The higher the federal funds rate, the more expensive it is to borrow money. Since it is only applicable to very creditworthy institutions for extremely short-term (overnight) loans, the federal funds rate can be viewed as the base rate that determines the level of all other interest rates in the U.S. economy. \\

The Federal Open Market Committee (FOMC), which is the Federal Reserve’s primary monetary policymaking body, telegraphs its desired target for the federal funds rate through open market operations. A rise in the Federal Funds Rate indicates the FOMC attempting to curb burgeoning economic growth to prevent an inflationary period, and therefore the U.S. Dollar should strengthen. 

\subsection{10-Year Treasury}

The 10-Year Treasury Note Yield at Constant Maturity was also used. The Treasury Yield is the return on investment, expressed as a percentage, on the U.S. government's debt obligations (bonds, notes and bills). From another perspective, it is the interest rate the U.S. government pays to borrow money for different lengths of time. \\

The 10-Year Treasury in particular tends to signal investor confidence. When confidence is high, the ten-year treasury bond's price drops and yields rise because investors feel they can find higher returning investments and do not feel the need to play it safe. But when confidence is low, the price increases and yields fall as there is more demand for safe investment. Therefore, the higher the yield on long-term bonds like the 10-Year Treasury, the better the economic outlook, and the stronger the U.S. dollar. 

\section{Macroeconomic Forecasts}

Each of the indicators were forecast in order to generate a trading signal. A Support Vector Machine was utilized for this purpose. Support Vector Machines, or SVMs, are supervised learning models with associated learning algorithms that analyze data used for classification and regression analysis. Given a set of training examples, each marked as belonging to one of multiple categories, an SVM training algorithm builds a model that assigns new examples to one of the categories, making it a multiclass non-probabilistic linear classifier. An SVM model is a representation of the examples as points in space, mapped so that the examples of the separate categories are divided by a clear gap that is as wide as possible. New examples are then mapped into that same space and predicted to belong to a category based on which side of the gap they fall. \\

Support Vector Machines have been found to particularly show excellent performance in the field of time series prediction. \cite{svm} SVMs can accurately  forecast time series data when the underlying system processes are typically nonlinear, nonstationary and not defined apriori. \\

However, forecasting the exact value that the macroeconomic indicators would hold on a month-to-month basis is a infeasible problem, even for SVMs. A vast amount of feature engineering would need to be done, which is both difficult and expensive. Variables would need to be created using domain  knowledge of the data to give the SVM the adequate data needed to make accurate forecasts. But, for our purposes, forecasting the direction in which the macroeconomic indicators move would be sufficient, and this is certainly a easier problem to solve. \\

An SVM was trained on monthly data, acquired through the Haver Analytics Database, from 1995-2008, in order to successfully forecast whether the indicator of note increased, decreased, or stayed constant in the evaluation period, 2008-2014, on a month-by-month basis. With little feature engineering, a classification accuracy of 70\% was yielded. \\

With these monthly forecasts, a trading signal can be generated. If the forecast details that the indicator will stay constant, a "hold" position was taken. If instead it predicts that the indicator will increase, then a "sell/short" position was taken, and if a decrease is predicted, a "buy/long" position was taken. This is because if, for example, the indicator were to increase, the value of the major currencies relative to the U.S. Dollar would decrease, meaning a "sell/short" position should be taken. \\

These monthly signals were normalized to daily signals, in order to be compared with the mean reverting signal. 

\section{Signal Weighting}

In order to combine the 4 signals into one, the signals were one-hot encoded. If each of the possible signals were encoded with nominal values, the ordinal property would cause greater values to have greater weight. Instead, a boolean vector is generated for each possible position, and only one of these vectors can take on the value "1" for each sample. \\

With the signals one-hot encoded, each possible signal now mathematically is equal in value. With that, a weight vector was generated which signifies how much weight will be given to each indicator's signal. The weights were bounded on the range 0 to 1, and these weights were then optimized using Sequential Least-Squares Programming (SLSQP), an iterative method for nonlinear optimization. The objective function here was the APR, and the near negligible transaction costs were not considered for the optimization. \\

The results are summarized in the following table: 

\begin{table}[h]
\centering
\begin{tabular}{l l l l l}
\hline
10-Year & S\&P 500 & Federal Funds & Mean Reversion & \textbf{APR}\\
\hline
0 & 0 & 0 & 1 & 4.084\% \\
0.5 & 0.5 & 0.75 & 1 & 4.112\% \\
\hline
\end{tabular}
\caption{Optimization Results}
\end{table}

If only the mean reverting signal was considered, an APR of 4.084\% was yielded. After optimization, giving non-zero weight to the macroeconomic indicators was found to be ideal, with a increased APR of 4.112\% obtained. With an evaluation period spanning 6 years and only 3 macroeconomic signals stemming from satisfactory forecasts considered, this is certainly a significant improvement. \\

\begin{figure}[h]
\centering\includegraphics[width=\textwidth,height=\textheight,keepaspectratio]{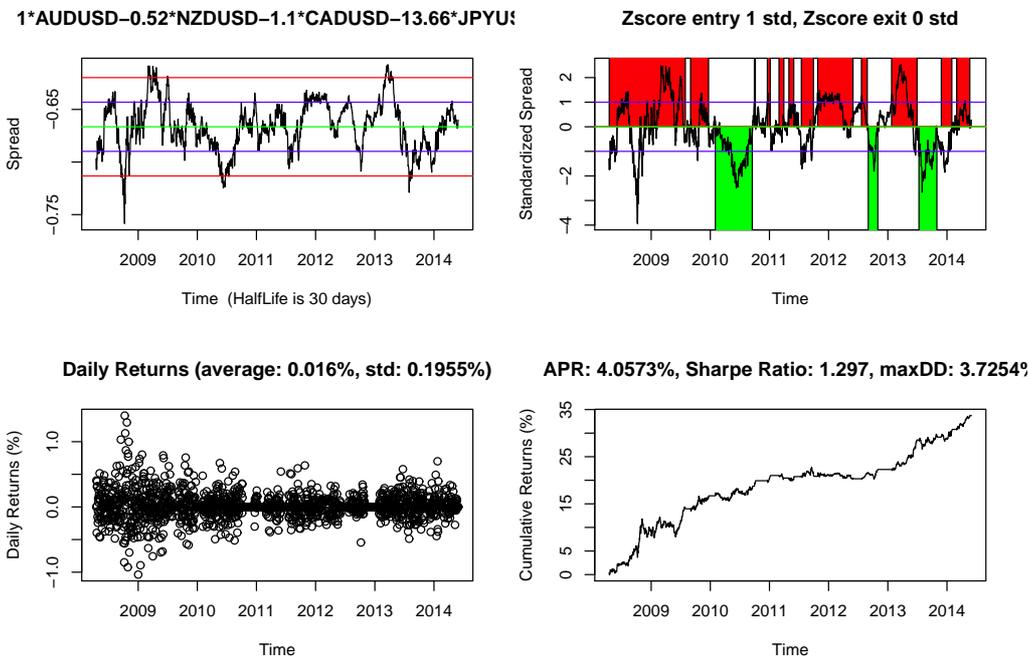}
\caption{Optimal Signal Weighting Results}
\label{fig:macroresults}
\end{figure}

The full backtested results can be seen in Figure~\ref{fig:macroresults}, and can be compared to the results generated using solely the mean reverting signal, summarized in Figure~\ref{fig:MeanReversionResults}. The lessening in the APR is due to the transaction costs factored in. Qualitatively evaluating the positions taken by examining the plot on the upper-right reveals the benefit to giving non-zero weight to the macroeconomic signals. \\

Comparing the historical positions taken reveals a similar pattern, except for the 2008-2010 period. The recession in the United States spanned the majority of that time, beginning December 2007 and ending June 2009. During this time, rather than continuing to rapidly transition between taking short and long positions based solely on the level of the spread, the macroeconomic indicators influenced the strategy to simply hold a short position throughout that time, betting that though the U.S. dollar was weak during this recessionary phase, it should revert back to the mean. That decision was the significant difference between the strategies, and what primarily net the improvement in performance.

\newpage

\section{Conclusion}
The Yield Curve is a strong economic indicator, and studies suggest that trading upon the mean-reverting nature of its spread yields statistically significant positive performance results. Due to that, a Multiple Pairs Trading Strategy was implemented using major currency pairs. Using the Johansen test, stationary cointegrated portfolios were yielded from the major currency pairs considered. With that, a mean reversion strategy was executed, by computing z scores and comparing them to the entry and exit thresholds. The representative cointegrated portfolio yielded a APR of 4.084\%. By using a SVM to generate monthly forecasts on a set of macroeconomic indicators, trading signals for each of the indicators considered were obtained. After optimizing upon the weights for each of the indicators, it was found that giving the macroeconomic signals nonzero weights yielded an APR of 4.112\%, an improvement upon the original strategy. With a large evaluation period and a small indicator set, this improvement is significant and demonstrates that macroeconomic indicators can certainly improve trading strategies. \\  




\bibliographystyle{model1-num-names}
\bibliography{sample.bib}







\end{document}